\documentclass[aps,pre,preprint,superscriptaddress]{revtex4}
\usepackage{graphicx}% Include figure files
\usepackage{graphics}% Include figure files

\begin{document}

%%%%%%%%%%%%%%%%%%%%%%%%%%%%%%%%
%  Definitions to write chemical equations
%%%%%%%%%%%%%%%%%%%%%%%%%%%%%%%%
\def\longrightharpoonup{\relbar\joinrel\rightharpoonup}
\def\longleftharpoondown{\leftharpoondown\joinrel\relbar}

\def\longrightleftharpoons{
  \mathop{
    \vcenter{
      \hbox{
    \ooalign{
      \raise1pt\hbox{$\longrightharpoonup\joinrel$}\crcr
      \lower1pt\hbox{$\longleftharpoondown\joinrel$}
    }
      }
    }
  }
}

\newcommand{\rates}[2]{\displaystyle
  \mathrel{\longrightleftharpoons^{#1\mathstrut}_{#2}}}
%%%%%%%%%%%%%%%%%%%%%%%%%%%%%%%%%%
%%%%%%%%%%%%%%%%%%%%%%%%%%%%%%%%%%%

%\preprint{}

%\title{Model of DI particle virus-host dynamics to explain viral latency}
\title{Stochastic model of virus and defective interfering particle spread across mammalian cells with immune response}


\author{Neil R. Clark}
%\thanks{}
\affiliation{Department of Pharmacology and Systems Therapeutics, Mount Sinai School of Medicine, New York, NY 10029, USA}

\author{Karla A. Tapia}
%\thanks{}
\affiliation{Department of Pathobiology, University of Pennsylvania, Philadelphia, PA 19104, USA}

\author{Aditi Dandapani}
%\thanks{}
\affiliation{Department of Applied Mathematics, Columbia University,
New York, NY 10027, USA}

\author{Benjamin D. MacArthur}
%\thanks{}
\affiliation{School of Mathematics, University of Southampton, Southampton, SO17 1BJ, UK}

\author{Carolina Lopez}
%\thanks{}
\affiliation{Department of Pathobiology, University of Pennsylvania, Philadelphia, PA 19104, USA}

\author{Avi Ma`ayan}
\email[]{avi.maayan@mssm.edu}
%\thanks{}
\affiliation{Department of Pharmacology and Systems Therapeutics, Mount Sinai School of Medicine, New York, NY 10029, USA}


\date{\today}

\begin{abstract}
Much of the work on modeling the spread of viral infections utilized
partial differential equations. Traveling-wave solutions to these
PDEs are typically concentrated on velocities and their dependence
on the various parameters. Most of the investigations into the
dynamical interaction of virus and defective interfering particles
(DIP), which are incomplete forms of the virus that replicate
through co-infection, have followed the same lines. In this work we
present an agent based model of viral infection with consideration
of DIP and the negative feedback loop introduced by interferon
production as part of the host innate immune response. The model is
based on high resolution microscopic images of plaques of dead cells
we took from mammalian cells infected with Sendai virus with low and
high DIP content. In order to investigate the effects discrete
stochastic microscopic mechanisms have on the macroscopic growth of
viral plaques, we generate an agent-based model of viral infection.
The two main aims of this work are to: (i) investigate the effects
of discrete microscopic randomness on the macroscopic growth of
viral plaques; and (ii) examine the dynamic interactions between the
full length virus, DIP and interferon, and interpret what may be the
evolutionary role of DIP. We find that we can explain the
qualitative differences between our stochastic model and
deterministic models in terms of the fractal geometry of the
resulting plaques, and that DIP have a delaying effect while the
interaction between interferon and DIP has a slowing effect on the
growth of viral plaques, potentially contributing to viral latency.

\end{abstract}

%\pacs{89.75.Fb, 05.10.Gg, 07.05.Tp}
%\keywords{}

\maketitle

\section{Introduction}

% Introduce DIP
% preocupation with growth rates and velocity

%%%%%%%%%%%%%%%%%%%%%%%%%%%%%%%%%Need to begin introduction with modeling of viral infections.

% p210 Population dynamics
% p437 Population dispersal Skellam 1951

The modeling of the dynamics of viral infection across host cells is
a classical problem in the field of population dynamics and
dispersal. Partial differential equations (PDEs) models of such
systems have a long history; notably Skellam
\cite{skellam1951random} was the first to apply PDEs to the random
dispersal of biological populations. Such models apply the continuum
assumption whereby populations of individuals are represented as
scalar concentration fields which obey PDEs. Many of these models
are not analytically solvable, and generally, simple solutions such
as traveling wave, are commonly used to describe the dispersal of
virus across host cells. Here we study the spatial and stochastic
effects of the dispersal of virus amongst an immobile space of host
cells with the use of an agent-based model. In this context we
examine the dynamic interactions of the virus with defective
interfering particles (DIP), which are incomplete forms of the
virus, described in detail below, and the interferon production by
host cells' immune response.

The system analyze here is made of a continuous monolayer of host
cells and a distribution of full-length and defective viral
particles as well as interferon molecules. The virus spreads by
infecting cells, replicating, then releasing it's yield upon killing
the host cell; this yield of virus particles are then free to
diffuse and infect neighboring cells, generating a growing plaque of
dead cells. The host cells' immune response can detect defective
viral particles and this results in the release interferon molecules
that locally reduce the probability of further viral infection. The
addition of the negative feedback loop through the interferon
response by host cells due to DIP detection was not previously
modeled by others.

A typical approach to model the dispersal of a virus across host
cells is to use the continuum assumption whereby the distribution of
particles and cells are represented as scalar concentration fields
which are solutions to differential equations. A system of
differential equations which embodies the hypothesized significant
mechanisms is derived and studied for insight into the population
dynamics. This approach has been popular in the field of immunology
modeling to enhance understanding of HIV-1 infection and other
pathogens \cite{Perelson02}, and to explore the idea that DIP could
be used for HIV therapy \cite{Nelson95}, while in
\cite{stauffer2010population} the authors investigated the
population dynamics of virus and DIP in serial passage cultures with
recurrence relations.

An extension of this approach is to study the spatio-temporal
dynamics of a population spread. The continuum assumption being made
even on the scale of whole organisms \cite{skellam1951random} being
among the first to do this. Typically partial differential equations
such as,
\begin{equation}
\frac{\partial P}{\partial t} = \mathcal{D}\nabla^2 P +\alpha P
\end{equation}
where $\mathcal{D}$ is the dispersal rate and $\alpha$ is the
intrinsic growth rate, form the basis of these models. One of the
first applications of this class of models to study virus-DIP
infection was by Frank \cite{Frank00}, however this model did not
give a full treatment of the spatio-temporal development and did not
include the immune response. A PDE model was developed by Yin and
McCaskill \cite{yin1992replication}, wherein the spread of the virus
was represented as a reaction diffusion system,
\begin{equation}
{\rm V} . {\rm H} \rates{k_{\rm 1}}{k_{\rm 2}} I  \displaystyle \mathop{\longrightarrow}^{\rm k_{\rm 2}}{\rm  Y} .{\rm  V}
\end{equation}
where $V$, $I$ and $H$ represent the concentration of virus,
infected host cells and uninfected host cells, and $k_1$, $k_{-1}$,
and $k_{2}$ represent the rates of viral infection, desorption and
cell death. The authors devised the corresponding PDE model, looked
for traveling wave solutions, and considered the dependence of the
velocity upon the parameters of the model. A similar approach was
taken in \cite{you1999amplification}. Haseltine
\cite{haseltine2008image} took this approach one step further by
fitting their model to images of growing viral plaques, while also
concentrating on the velocity. More recently Amore \cite{
amor2010virus} expanded upon reaction-diffusion models by including
the delay time between infection of a cell and the release of viral
progeny. The preoccupation with velocity permeates most of the
literature on this subject, however here we are mainly concerned
with the qualitative details of the spatio-temporal dynamics of
viral infections, it's implications for the dynamical significance
of the stochastic nature of the physical and biological mechanisms,
focusing on the possible role for DIP in the context of their
detection by the host and the resultant interferon immune response.

Here, rather than velocity, we investigate the qualitative dynamic
effects of the mechanisms of virus dispersal, adding the important
variables of interferon and DIP. DIP were discovered in the 1950's
as incomplete forms of the influenza virus that interfere with viral
replication \cite{VONMAGNUS54, Paucker58}. They were subsequently
observed for almost all RNA viruses such as rabies \cite{Clark81},
sendai (SeV) \cite{Kolakofsky76}, polio \cite{Cole71}, sindbis
\cite{Shenk73}, vesicular stomatitis virus (VSV) \cite{Schubert78},
and measles \cite{Hall74, Rima77}. It was discovered that DIP
interfere with viral replication by over loading the viral
replication machinery because shorter DIP replicate faster compared
with the production of full-length virus \cite{Yount06}. It was also
discovered that DIP can be detected by the host, promoting
interferon production leading to a robust immune response
\cite{Marcus77, Lopez06, Yount06}. Many DIP can only replicate
through co-infection with the full virus leading to a parasitic or
predator-pray type relationship.

DIP are a conserved biological phenomenon with no known function.
Given that many species of virus are often found to co-exist with
their corresponding DIP, it is reasonable to suppose that they could
be performing a biological function that confer an evolutionary
advantage, or otherwise exist in some kind of evolutionary
equilibrium with the virus. However, this is not currently known or
proposed, and the above properties, being ostensibly detrimental to
the virus, do not signal an obvious function or mutual evolutionary
advantage. Here we shall consider, through a generic model, the
dynamical interaction between virus, DIP, host cells and their
innate immune response. We demonstrate that DIP can have a delaying
effect on the spread of virus, and interferon can have a slowing
effect. We provide some insight into these specific relationships.
Our model leads us to a more general consideration of the continuum
assumption behind PDE approaches to the modeling of virus spread in
an immobile space of host cells. The most significant result of this
part of our investigation is that, in this system, a discrete
stochastic model may have qualitatively different solutions than the
deterministic, traveling-wave solutions of reaction-diffusion PDE
models. In which case it is important to determine which type of
model is most appropriate for this biological system, and this
determination may extend understanding of the most important
mechanisms in the spread of viral infection in a host.

Our model is primarily based on high-resolution microscopy images we
took of stationary primate kidney cell line in culture infected with
Sendai virus (SeV) with or without DIP. We construct a stochastic
agent based model which does not rely on the continuum assumption,
and retains the discrete and random nature of viral infection and
decay. We explore the qualitative properties of the solutions for
various values of the parameters. By comparing the output of our
agent based model to PDE models, we observe that the stochastic
model of plaques are growing with an accelerating speed. The
mechanism by which this occurs is explored quantitatively in terms
of the fractal geometry of the model plaques. The dynamic effect of
DIP and the interferon response is gauged qualitatively, and it is
found that DIP can delay the growth of viral plaques while
interferon can slow their growth. Hence, the known biological
properties of DIP could potentially explain their moderating effect
on viral plaque growth.

\subsection{Biological and experimental background}
We aim to model the spread of an RNA virus, its DIP and the
interferon response of the immune system, through a monolayer of
living mammalian cells in a dish. As the virus spreads, a region of
dead cells is formed called a {\it plaque}. Most of our results will
concern the properties of these plaques. The construction of the
model is based on an abstraction of the mechanisms of virus spread
which we based on microscopic images we collected experimentally
from stained plaques. In this experiment LLCMK2 cells were grown in
Dulbecco's modified Eagle's medium (Gibco-BRL) supplemented with
10\% fetal calf serum (FCS; Gibco-BRL), Sodium pyruvate, L glutamine
and gentamicin. The cells were plated in 24 well plates. Confluent
monolayers were infected with 200 infectious particles of SeV (low
DIP) alone or SeV together with 2000 DIP, or mock infected. After 1h
incubation at 37C the cells were overlaid with 500ul of agar melted
in infection media containing 0.025 mg of trypsin (Worthington). The
infected LLCMK2 cells were then fixed with 4\% paraformaldehyde at
48 or 96 hrs post infection and blocked overnight at 4C with PBS/BSA
1\%. Cells were then stained with a monoclonal anti-SeV NP antibody
(clone 3F11) for 45 min at room temperature, washed twice with
PBS/BSA 1\%, and incubated with a peroxidase-conjugated secondary
antibody (Jackson ImmunoResearch) for 45 min at room temperature.
After washing, the staining was developed using the AEC Substrate
kit according to manufacturer's recommendations (BD Pharmingen).
Picture scanning of the wells (10x) was taken using a Zeiss
Axioplan2IE microscope and montage stitching done with the Metamorph
software (MDS Analytical Technologies) at the MSSM-Microscope Share
Resource Facility (Fig.\ref{fig:photo}). This figure shows the
roughly circular outlines of the monolayer of cells, where those
stained red have been killed by the virus. We can immediately see
from these images that DIP appear to have arrested the growth of
viral plaques.

The abstraction of the mechanisms of spread of infection that we
envisage is that initial infection is nucleated, being seeded by the
infection of an individual cell with an individual virus particle.
The virus replicates internally for some time before the cell is
killed and the virus yield is released. The released viral particles
then diffuse freely until they either decay or they infect a healthy
neighboring cell and thereby spread the infection. This abstraction
is consistent with the observation of distinct plaques of dead cells
which grow by an expansion of their boundaries, which are quite
sharp and irregular. A low virus yield ($O(10)$ particles) and
significant decay rate ($ O(10^{-5} \text{s}^{-1})$ ), and the
discrete random nature of infection, replication and decay, suggests
that stochastic effects may be significant at the intercellular
scale. In this case we would not make the continuum assumption.
However, the local growth of even macroscopic viral plaques is
expected to be generated by the same microscopic mechanisms as the
intercellular spread. We investigate the stochastic effects at the
intercellular scale on the macroscopic growth of viral plaques by
developing a model which incorporates them explicitly. Furthermore
we examine the qualitative effects of DIP and the interferon immune
response on the dynamics of viral infection in our model.

\subsection{Model construction}
The model is agent-based, such that the hypothesized significant
biological entities are represented as agents, and the significant
biological and physical mechanisms are represented as the rules of
behavior of the agents. The model results in stochastic simulations
of individual viral plaques in which each individual viral particle,
interferon molecule and cells is retained explicitly. We begin by
describing the agents, their states and their corresponding
biological entities before describing the rules and the parameters.

The monolayer of cells is represented by a square lattice. Each
element of the lattice can be in two states corresponding to a
living or dead cell. This lattice also serves as the discretization
of space in which the viral particles, DIP and interferon molecules
are located. The total number of each type of particle at each
lattice point at each time is stored. The viral particles can be in
two states, internal and external. In the external state the
particles are free to diffuse on the lattice, however, internal
particles correspond to those which are residing inside living
cells. Agents representing interferon molecules are secreted by
living cells and are never internal in the model. Figure
\ref{fig:agents} shows a schematic description of these agents.

The rules of behavior of the agents encode the following biological
and physical mechanisms: diffusion, infection, decay, replication,
interferon secretion and cell killing. First, diffusion is
incorporated by making free agents perform a discrete random walk on
the lattice, the timestep is chosen to be consistent with the
prescribed diffusion coefficient. At each discrete timestep viral
particles undergo an independent Bernouli trial which determines
weather they decay or, if the cell at it's current lattice position
is alive, the viral particle may infect the cell and thereby become
internal. The probabilities of the Bernouli trials are set to be
consistent with the prescribed viral infection and decay rates.
Internal viral particles replicate via a Poisson process with rates
consistent with the prescribed yield and lifetime of an infected
cell. The replication obeys the known logic of the interaction of
virus and DIP such that (i) virus alone - replicates at rate r (ii)
DIP alone - no replication (ii) virus and DIP confection - virus
replicates at rate r/$\rho$, DIP replicate at rate r, where $\rho
\approx O(20)$. This parameter has experimental backing in from
Yount et al. \cite{Yount06}. The effect of this parameter is not
investigated here since it has experimental backing and exploring
its effect on the model is beyond the scope of this study. Cells
infected with DIP can detect these particles, and this leads to
secretion of interferon molecules at a prescribed rate, treated as a
free parameter. The effect of the interferon concentration is to
locally reduce the probability of viral infection. We model this
with a hill-function:

\begin{equation}\label{eqn:probinf}
p_i=\frac{p_{i,0}}{1+\beta I}
\end{equation}
Where $p_{i,0}$ and $\beta$ are constant parameters and $I$ is the
concentration of interferon. Finally, when the virus/DIP has
replicated up to it's yield, the cell dies and the internal
particles become external.

The model parameters dictate the spatial dimensions of the lattice,
the diffusion coefficients, and the rates of the various processes.
We set the grid spacing to be equal to the approximate cell spacing
in our experimental monolayer, $~ 20 \mu m$, and the grid size to
400 cells$^2$, so that we can investigate macroscopic plaques. A
base set of parameters, shown in table \ref{tab:baseparameters}, is
chosen consistent with You and Yin \cite{you1999amplification}, then
perturbations around this set are made in runs of the model. The
initial condition is for a single infected cell in the center of the
lattice which seeds the growth of an individual plaque. The lattice
has periodic boundary conditions however the model is not run for
enough time for the effects of the boundary to have an effect on the
results.

\section{Results}

Figure \ref{fig:modelplaque} shows the development of an individual
model plaque: the spatial distribution of the dead cells, the free
virus, DIP and interferon at several times. The model plaques have a
compact morphology with irregular boundaries. The free virus and DIP
reside predominantly on the periphery of the plaque, and so the
plaque grows by an expansion of it's boundaries. Most previous
approaches to this type of problem have been concerned with the
velocity of traveling wave solutions. However, here we are concerned
with the qualitative solutions to our model and their difference
from traveling waves. Later we consider the qualitative effects of
DIP and interferon.

First, we examine the growth of the number of dead cells in a plaque
for various parameter sets. Figure \ref{fig:growthrates} shows the
number of dead cells in model plaques against time, where the gray
and black curves correspond to runs in which interferon was present
and left out of the model respectively. We observe an initial phase
of fast growth, where the plaque is  $O(10)$ cells in number. This
fast growth is dominated by nucleation events. There are few killed
cells in the vicinity, and hence the growth is not significantly
limited by the presence of dead cells.

At later times, when the plaque is  $O(100)$ cells in number, the
growth curves change to a power-law, when no-interferon is
introduced in the model. Growth with a monotonically reducing
exponent in the presence of interferon is observed. The line in the
figure indicates the slope of a power law quadratic in time; we can
see that the model plaques are growing faster than this rate. In the
following section we concentrate on the power-law growth. A
power-law growth is perhaps not surprising for a diffusion-limited
growth such as the model we present here. However, the exponent of
the power-law is of interest to us because it involves a qualitative
difference to the traveling wave solutions. Subsequently we shall
address the effect of the DIP and interferon on the overall behavior
of the model.

\subsection{Accelerating plaque growth}
In the absence of interferon, the growth curves of model plaques
shown in figure \ref{fig:growthrates} obey a power-law with an
exponent greater than $2$ which means that their mean radius
accelerates their expansion. A plaque with a simple geometry growing
by linear expansion, as is the case for a traveling wave, would grow
as $t^2$, where $t$ is time. We aim to interpret this difference in
terms of the geometry of our model plaques with the aim of
illuminating the importance of the microscopic stochasticity for the
qualitative nature of the growth of macroscopic viral plaques. We
aim to do this with a phenomenological argument.

Part of the abstraction for the mechanism of plaque growth is that
the virus resides on the periphery of the plaque and the plaques
grow when the infected cells on the periphery die and release viral
particles to infect nearby cells. One explanation for the observed
qualitative difference in the observed exponent of the plaque
power-law growth is that the irregular geometry of the plaques gives
them a larger perimeter for a given area and therefor more infected
cells for a given number of dead cells, resulting in a greater
exponent in the power-law growth.

This would rely on a fractal plaque geometry, where the fluctuations
in the plaque boundary are scale-free, and increase in their range
of scales as the plaque develops; in which case the plaque would
have the following fractal area-perimeter (A-p) relationship:
\begin{equation}\label{eqn:aprelationship}
A\propto p^\frac{2}{D}
\end{equation}
where $D$ is the fractal dimension of the plaque boundary.

We begin by demonstrating the fractal nature of the model plaques.
Figure \ref{fig:aprelationship} shows the area-perimeter
relationship for model plaques in the absence of interferon. We see
that in each case the exponent is greater than $2$, indicating a
fractal dimension of the plaque boundaries which is greater than
unity. In order to further demonstrate the scaling nature of the
plaque boundary fluctuations, we calculate the radial coordinates,
origin at the center of the lattice, of infected cells on the
boundary, and plot the radial against the angular component in a
Cartesian plot \ref{fig:polarplot} at three different times in the
development of an individual plaque. In each case there are
fluctuations at the spatial scale of the lattice spacing. However,
we can see that the range of scales increases with time as the
fluctuation curve extends over an ever larger range of scales. It
may appear from the images of the plaques that they become more
circular with time, this is because the range of fluctuations scaled
with the mean radius of the plaque decays. However, the absolute
scale of the fluctuations increases. This is illustrated in figure
\ref{fig:scaling} which shows the root-mean-squared fluctuation
intensity as a function of time for an individual representative
model plaque.

Next we consider the relationship between the number of infected
cells, $n_{i}$ and the plaque perimeter $p$. Direct proportionality
is unlikely because there is a band of infected cells of finite
width which follows the perimeter, so we would expect fluctuations
in the perimeter of the order of this width to be smoothed out, such
that
\begin{equation}
n_{i} \propto p^{\gamma}.
\end{equation}
But because we observe this band to be thin, we may expect an
exponent, $\gamma$, close to but smaller than unity. In figure
\ref{fig:nprelationship} we plot the number of infected cells
against the plaque perimeter for all model plaques which developed
in the absence of interferon. We observe an exponent which is
marginally less than unity in each case.

To test the hypothesis that the fractal geometry of our plaques
contributes to their accelerating growth, we analyze the expected
exponent of the growth given the fractal dimension of the plaque and
the relationship between the number of infected cells and the
perimeter.

If we suppose that the rate of change of the number of dead cells in
a plaque, $Q$, is in proportion to the number of infected cells,
$n_{i}$, ignoring the time delay between infection and cell death,
\begin{equation}
\frac{dQ(t)}{dt}\propto n_i
\end{equation}
and we take the relationships that follow from the fractal geometry
of the plaques,
\begin{equation}
n_{i}\propto p^{\gamma}
\end{equation}
along with equation \ref{eqn:aprelationship}, then if we take $Q$ to
simply be proportional to the plaque area we can write,
\begin{equation}
\frac{dQ(t)}{dt}\propto Q(t)^{\frac{\gamma D}{2}},
\end{equation}
which we can integrate to obtain,
\begin{equation}\label{eqn:predictgrowth}
Q(t) \propto t^{\frac{2}{2-\gamma D}}.
\end{equation}
For each model plaque, in the absence of interferon, we plot the
number of infected cells against the perimeter, the number of dead
cells against the perimeter, and the number of dead cells against
time (see supplementary figure \ref{fig:prodictgrowthdata}) and
least-squares fit to estimate the exponents $D$ and $\beta$, and
compare the predicted growth exponent from equation
\ref{eqn:predictgrowth}. This comparison is shown in figure
\ref{fig:predictgrowth}, where we can see that the degree of
correlation is partially limited by the errors in the estimation of
the exponents $D$ and $\gamma$. However, it is enough to suggest
that the fractal nature of the plaque geometry can at least
partially account for the exponent of the plaque growth curve.

\begin{figure}\label{fig:planefront}
 \centerline{\includegraphics[width=6in]{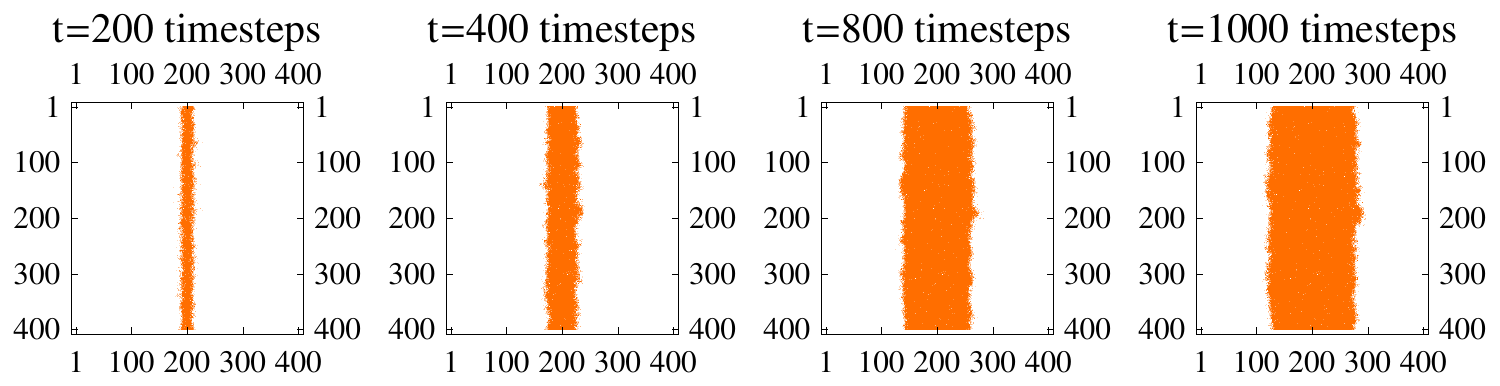}}
 \caption{A depiction of the distribution of dead cells in the spreading of a plaque which was started with the initial condition where a line of cells from the top to the bottom of the lattice are infected with virus. \label{fig:modelplaque}}
 \end{figure}

\begin{figure}\label{fig:planefronacc}
 \centerline{\includegraphics[width=3.375in]{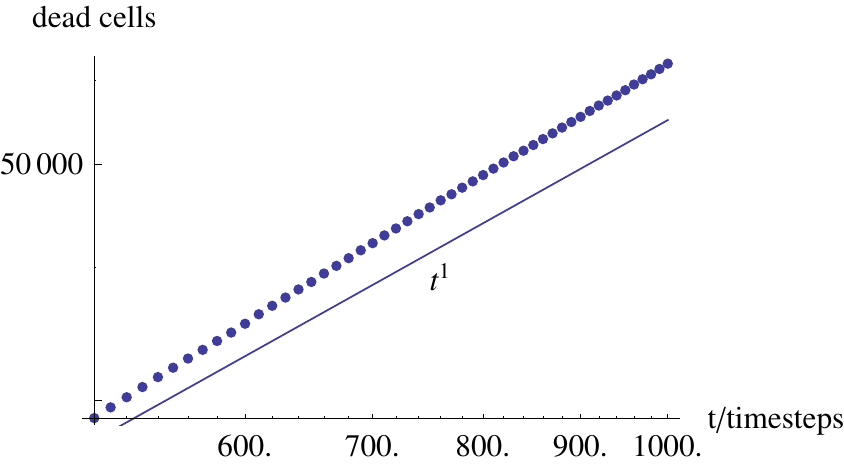}}
 \caption{The number of dead cells in the growth of a plaque.\label{fig:planefrontacc}}
 \end{figure}

Another potential explanation for the accelerating velocity of
plaque growth is mean curvature effects. As the plaque grows the
curvature of the mean boundary falls - it the plaque growth velocity
depended on this curvature then this could potentially account for
the acceleration. To test this we ran the model with a different
initial condition: a line of cells from the top of the lattice to
the bottom were set to be infected intimal; the resulting plaque was
a plane front which spread to the outer edges of the lattice (see
figure \ref{fig:planefront}). The mean plane front has constant
curvature, i.e., zero. Hence, any acceleration cannot be due to
curvature of the mean front. Figure \ref{fig:planefronacc} shows the
accelerating growth of this plaque, which appears to be due to the
development of fluctuations in its boundary geometry rather than the
mean curvature.

\section{The effect of DIP and interferon}
In our wet-lab experiment DIP outnumbered virus particles by about
an order of magnitude in order to generate a significant effect of
the infection. In the light of this, and the fact that DIP cannot
replicate alone, in order to simulate the effects of DIP on an
individual plaque we set the initial conditions such that the plaque
develops in a randomly uniform distribution of internal DIP. In
order to isolate the effect of the DIP, we examined the growth of
plaques with various concentrations of DIP in the absence of
interferon. Figure \ref{fig:dipeffect} shows the rate of growth of
plaques where the same parameters are used (see supplementary
figures) except $5\%$, $20\%$, and $30\%$ DIP. We observe that the
strongest effect of DIP appears to be a delaying effect on the
growth of the plaques such that the $30\%$ growth curve remains
about a factor of two smaller than the $5\%$.

In order to further investigate the effect of DIP, we plot the
effect of the concentration of DIP upon the exponent of the plaque
growth curve, the fractal dimension and the exponent of the
power-law relation between the number of infected cells and the
perimeter (see figure \ref{fig:dipeffect2}). We see that the fractal
dimension increases with DIP, and that the value of $\beta$ falls.
These two effects approximately cancel out such that the exponent of
the growth curve is not significantly affected. It appears that the
presence of DIP is felt in the early development of the plaque when
it is made of $O(100)$ cells in number, at that time the delay is
induced.

Figure \ref{fig:interferoneffect} shows model plaques with
parameters set with various interferon secretion rates, decay rates,
and strengths of effect on live cell (see equation
\ref{eqn:probinf}), while keeping all other parameters the same. The
aim in varying the interferon relevant parameters was to observe the
various qualitative changes to the model plaque growth curves due to
the dynamic effects of interferon. Broadly speaking the interferon
slows the growth of plaques from a power-law to a curve with a
monotonically decreasing exponent - possibly a modified logarithmic
growth. However, depending on the balance of the parameters
governing the strength of the effect of interferon on neighboring
cells, and the interferon decay rate, the plaques can display a
bi-phasic growth curve in which the curve is initially concave but
as the concentration of interferon reaches saturation, the curve can
return to a power-law form.

\section{Conclusions}
In the construction and execution of our model we have addressed two
main aims: (i) to investigate the qualitative differences in the
simulated plaque growth resulting from deterministic PDE models and
our discrete stochastic agent-based model; (ii) explore the
qualitative dynamic effects of DIP and interferon on the growth of
viral plaques.

We revealed that the agent-based model produces plaques which grow
faster than quadratically in time, this in contrast to most previous
work on such systems which look for traveling-wave solutions and
then focus on velocities. We found that the fractal geometry of the
plaques in the agent-based model can at least partly explain the
difference between the exponents.

The model indicated that DIP have a delaying effect on the growth of
model plaques, and that large amount of DIP relative to virus
particles are required to have an appreciable effect. DIP appear
hamper the growth of viral infections, as we observed that DIP
arrest the growth of our experimental SeV plaques. The model results
show that the impeded growth of viral plaques due to DIP can at
least partly be a dynamical effect which only depends on the known
biological properties of DIP. It is tempting to tentatively propose
the hypothesis that DIP dynamically impede the growth of viral
infection, and that this could be performing a useful function for
the virus in moderating its spread so it does not kill the host
before it is provided with the opportunity to jump host.

The final aim was to investigate the effects of the interferon
immune response. We found a range of qualitatively different growth
curves, the form of which depended critically on the parameters
governing the secretion of interferon and the protective effect of
interferon on uninfected cells. Broadly speaking, considering the
interferon response in the model slowed down the power-law growth to
one with a monotonically decreasing exponent, similar to a
logarithmic growth. As such, interferon has a much more dramatic
slowing effect on the growth rate of viral plaques.

\begin{acknowledgments}
We thank Drs. Charles Peskin from NYU, Jacob Yount from the
Rockefeller University, and James G. Wetmur from MSSM for useful
discussions. This work was supported by NIH grants 5P50GM071558-03,
1R01DK088541-01A1, and KL2RR029885-0109.
\end{acknowledgments}

\bibliography{DIP-library-bib}

\clearpage

\begin{table}[h]
\caption{Base set of model parameters, which is based on
\cite{you1999amplification}, and the decay rates are taken from
\cite{Frank00}.\label{tab:baseparameters}}

\begin{tabular}{@{\vrule height 10.5pt depth4pt  width0pt}lcc}
\noalign{\vskip 8pt}
%&\multicolumn5c{Repeat length}\\
\noalign{\vskip-11pt}
Parameter& Symbol&Value\\
%\cline{2-6}
%\vrule depth 6pt width 0pt years&\multicolumn1c{\it n}&Mean&SD&Range&Median\\
\hline

Virus infection rate & $k_{1, V}$  & $1.4 \times 10^{-10} \text{ml}/\text{hour}$ \\
DIP infection rate   &$k_{1, D}$  & $1.4 \times 10^{-10} \text{ml}/\text{hour}$ \\

Infected cell death rate & $k_{2,}$  & $5.91 \times 10^{-2} \text{hour}^{-1}$ \\

Virus decay rate & $k_{3, V}$ & $4.0 \times 10^{-5} \text{s}^{-1}$ \\
DIP decay rate & $k_{3, D}$ & $k_{3, V}$ \\

Interferon decay rate & $k_{3, D}$ & $k_{3, V}$ \\
Diffusion coefficient& $\mathcal{D}$ & $2.38 \times 10^{-6} \text{cm}^2/\text{hour}$ \\
lattice spacing& $dx$ &$20 \mu \text{m}$ \\
timestep & $dt$ & $\frac{dx^2}{4\mathcal{D}}=1.51\times 10^3 \text{s}$\\
Lifetime of infected cell & $L=k_{2}^{-1}$ &$17\text{hours}$\\
Yield & $Y$& 50 virus copies \\
rate of virus replication& $R_V$ & $Yield/Lifetime$\\
rate of virus replication& $R_D$ & $Yield/Lifetime$\\
rate of interferon  sectretion &S& rate of virus replication\\
Strength of interferon&$\alpha$ & 0.05 \\

\hline
\end{tabular}
\end{table}

\clearpage

 \begin{figure}\label{fig:photo}
\centerline{ \includegraphics[width=3.375in]{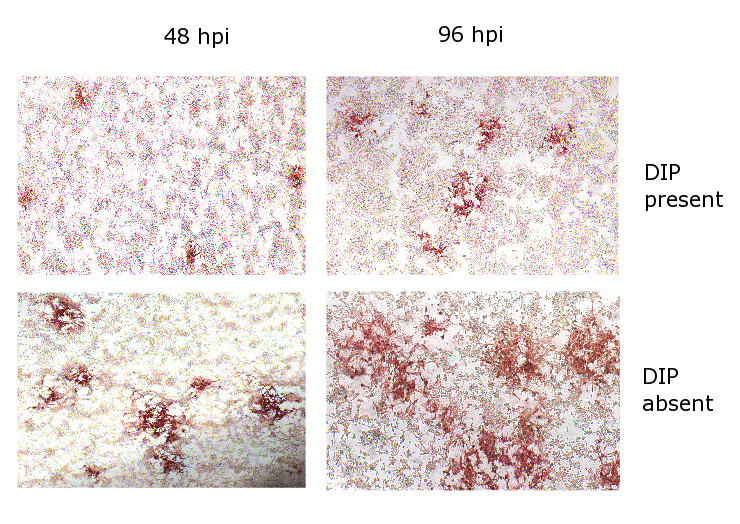}}
 \caption{Viral plaques grown \emph{in vitro}. High resolution images including mock treated cells are available from http://amp.pharm.mssm.edu/dip-high-res-images.zip. \label{fig:photo}}
 \end{figure}

 \begin{figure}\label{fig:agents}
 \includegraphics{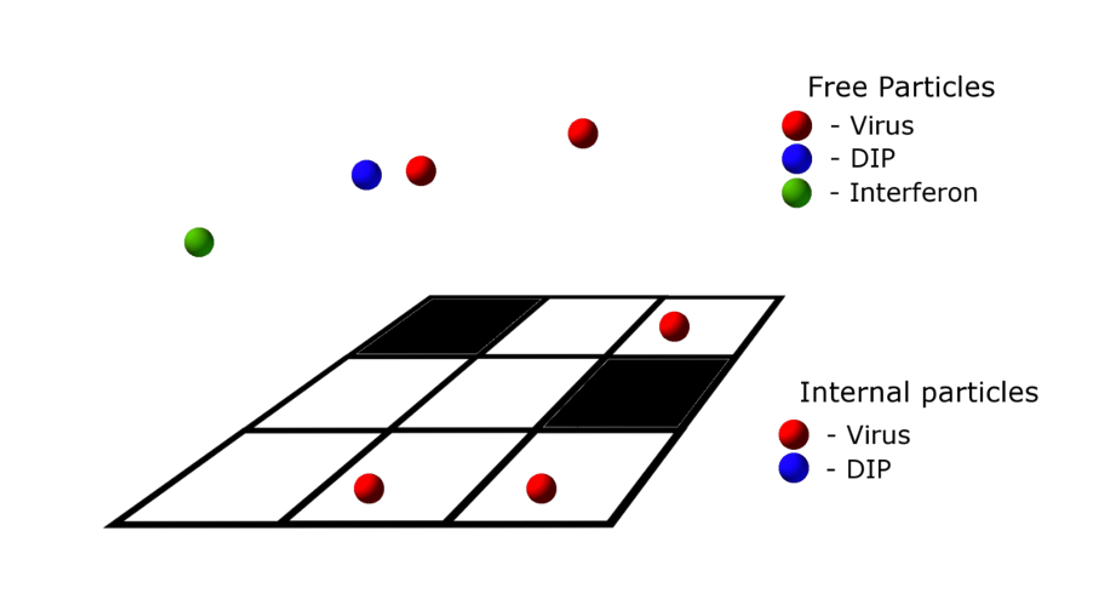}%

\centerline{ \includegraphics[width=3.375in]{agents}}
 \caption{A schematic illustration of the agents which represent the biological entities in the agent-based model.  \label{fig:agents}}
 \end{figure}

 \begin{figure}\label{fig:modelplaque}
 \centerline{\includegraphics[width=3.375in]{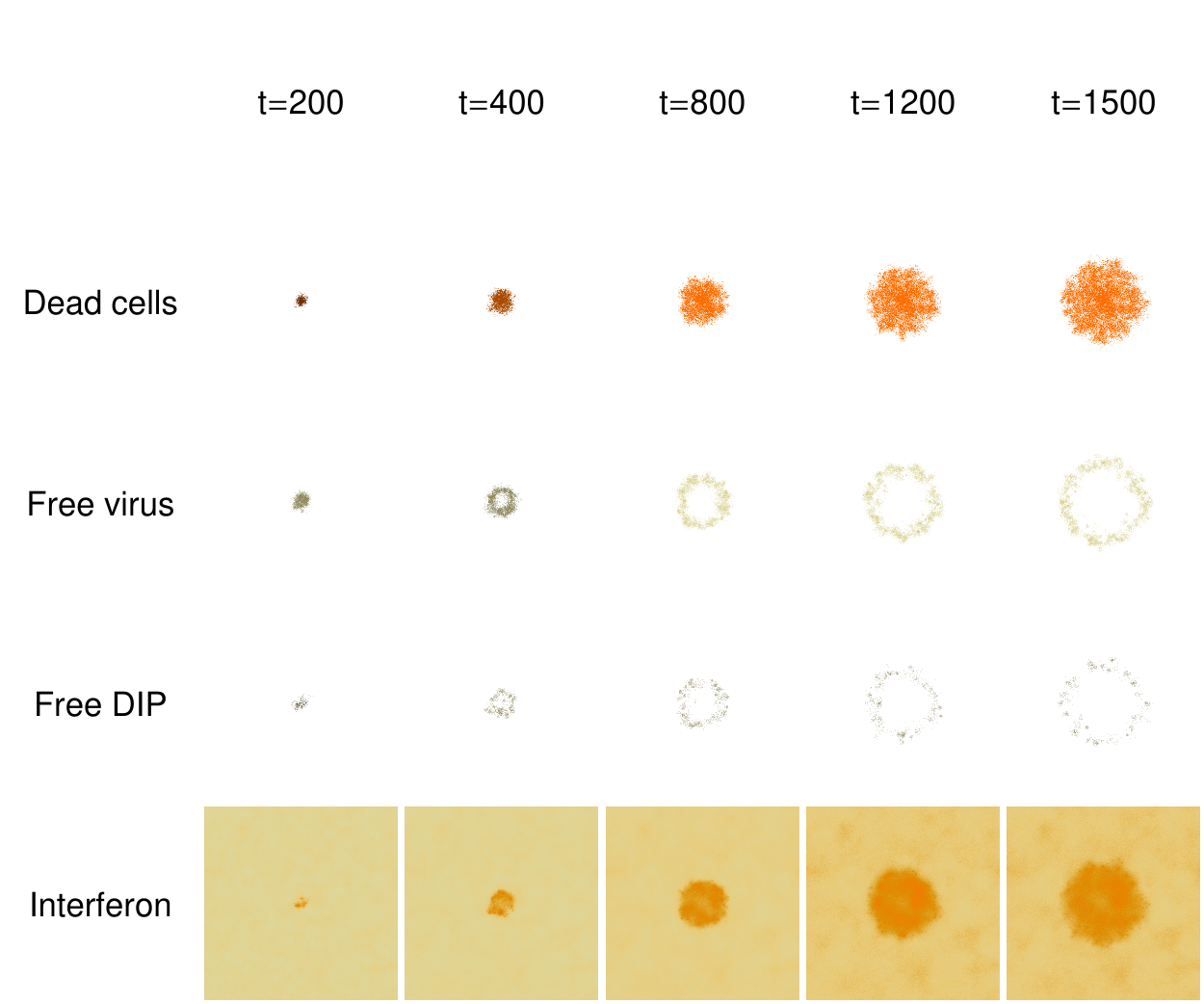}}
 \caption{A depiction of the distribution of killed cells, free virus, free DIP, and free interferon for a representative model plaque at four times.  \label{fig:modelplaque}}
 \end{figure}

%
% \begin{figure}
% \centerline{\includegraphics[width=3.5in]{dip-fig1a.pdf}}
% \caption{ }
% \end{figure}
%
% \begin{figure}
% \centerline{\includegraphics[width=3.375in]{dip-fig1b.pdf}}
% \caption{caption }
% \end{figure}
%
% \begin{figure}
% \centerline{\includegraphics[width=3.375in]{dip-fig1c.pdf}}
% \caption{caption}
% \end{figure}

\begin{figure}\label{fig:growthrates}
 \centerline{\includegraphics[width=3.375in]{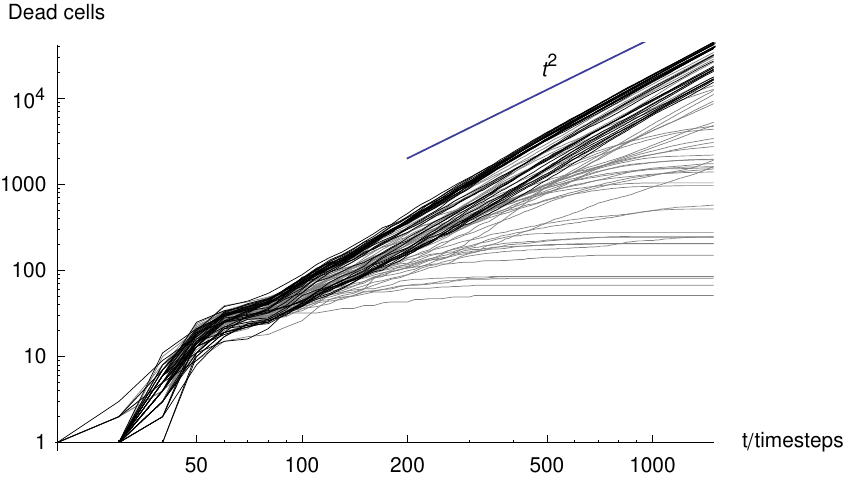}}
 \caption{The number of dead cells in individual plaques plotted against time. The black and gray curves correspond to plaques growing in the absence and presence of interferon respectively. The parameter sets for these growth curves can be found in the supplementary materials.  \label{fig:growthrates}}
 \end{figure}

\begin{figure}\label{fig:aprelationship}
 \centerline{\includegraphics[width=3.375in]{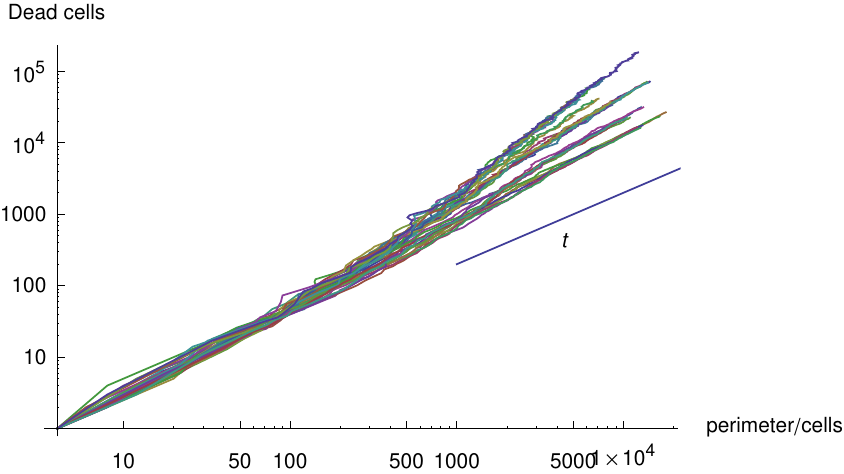}}
 \caption{ The number of dead cells plotted against the plaque perimeter. The number of dead cells is directly proportional the plaque area, so the scaling of this curve is the same as for the area-perimeter relationship and can be used to estimate the fractal dimension of the plaques. \label{fig:aprelationship}}
 \end{figure}

\begin{figure}\label{fig:polarplot}
 \centerline{\includegraphics[width=3.375in]{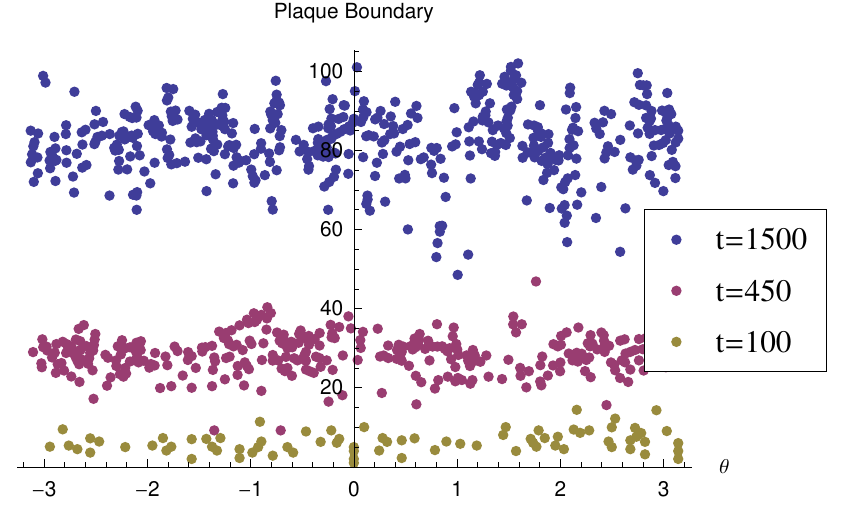}}
 \caption{The polar coordinated of infected cells residing on the plaque boundary for an individual plaque at three different times. The plot shows the increasing range of scales of the fluctuations in the shape of the model plaque boundary.  \label{fig:polarplot}}
 \end{figure}

\begin{figure}\label{fig:scaling}
 \centerline{\includegraphics[width=3.375in]{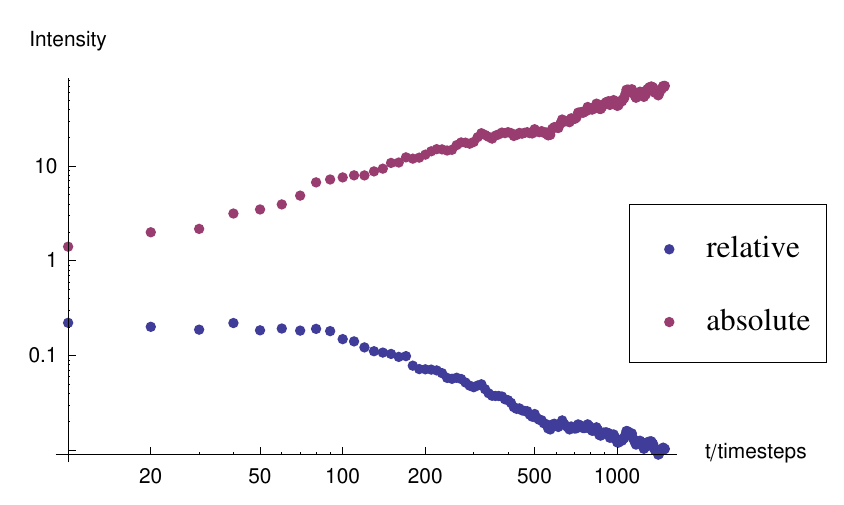}}
 \caption{The root-mean-squared intensity of the fluctuations in a representative, individual plaque boundary, both absolute, and scaled with respect to the mean radius of the plaque. The fluctuations decay with respect to the mean radius of the plaque, such that the plaques appear more circular and they grow larger, however the absolute intensity of the fluctuations increases as the model plaque grows, in a scale-free manner. \label{fig:scaling}}
 \end{figure}

 \begin{figure}\label{fig:nprelationship}
 \centerline{\includegraphics[width=3.375in]{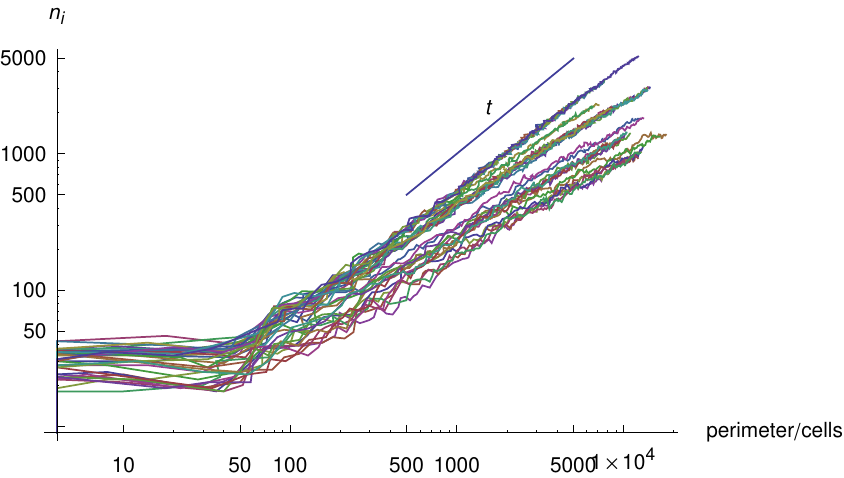}}
 \caption{The number of infected cells plotted against plaque the perimeter of individual plaques growing in the absence of interferon. \label{fig:nprelationship}}
 \end{figure}

\begin{figure}\label{fig:prodictgrowthdata}
 \centerline{\includegraphics[width=4in]{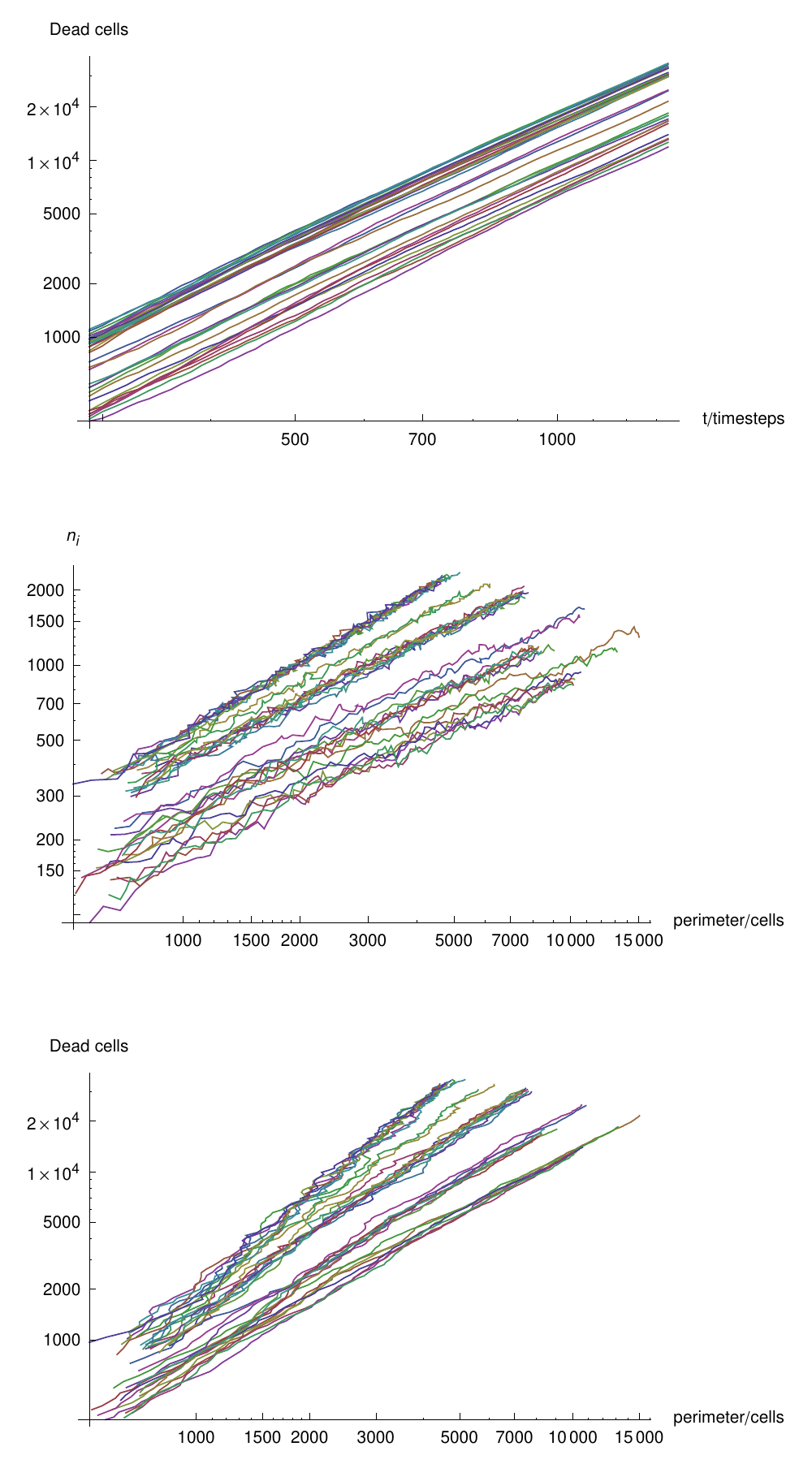}}
 \caption{The curves which are least-squares fitted and upon which the estimations plotted in figure \ref{fig:predictgrowth} are based.    \label{fig:prodictgrowthdata}}
 \end{figure}

 \begin{figure}\label{fig:predictgrowth}
 \centerline{\includegraphics[width=3.375in]{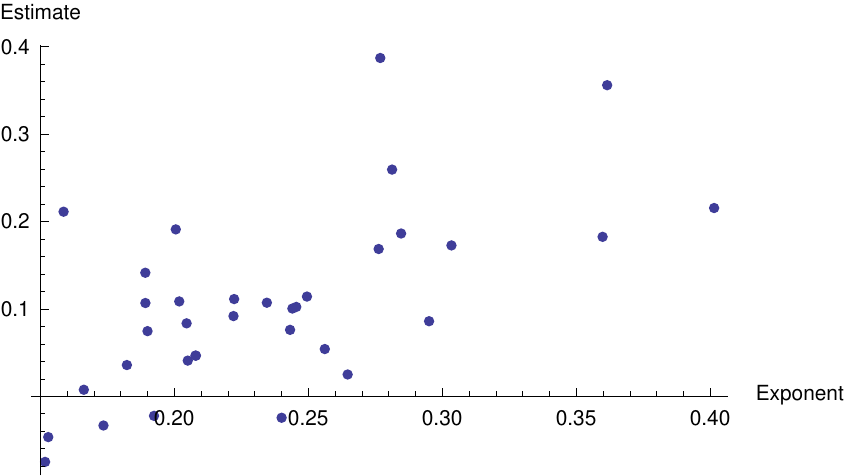}}
 \caption{The power-law exponent predicted from equation \ref{eqn:predictgrowth}, which is based on the fractal geometry of the plaques, plotted against the actual power-law exponent estimated by least-squares fitting. \label{fig:predictgrowth} }
 \end{figure}

 \begin{figure}\label{fig:dipeffect}
 \centerline{\includegraphics[width=3.375in]{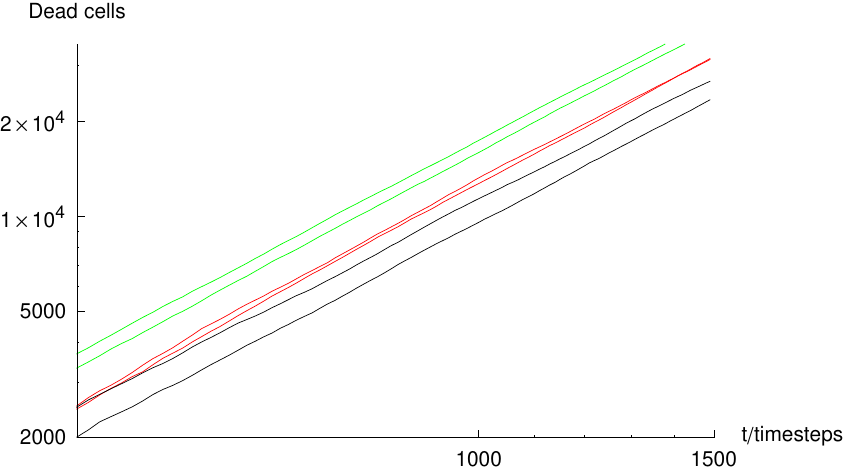}}
 \caption{The number of dead cells in plaques with identical parameter sets (see supplementary materials), but growing in the presence of different concentrations of DIP. The green, red and black curves correspond to $5\%$, $20\%$, and $30\%$ of cells infected with DIP respectively. \label{fig:dipeffect}}
 \end{figure}

 \begin{figure}\label{fig:dipeffect2}
 \centerline{\includegraphics[width=6in]{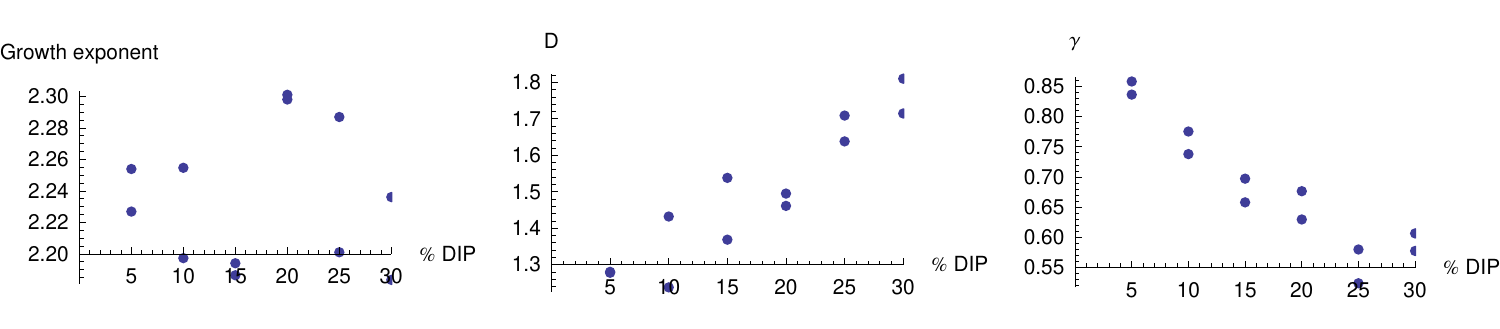}}
 \caption{The effect of the DIP concentration on the exponent in the plaque growth curve power-law, the fractal dimension of the plaque boundary, and the relation bet6ween the number of infected cells and the plaque perimeter. \label{fig:dipeffect2}}
 \end{figure}

 \begin{figure}\label{fig:interferoneffect}
 \centerline{\includegraphics[width=3.375in]{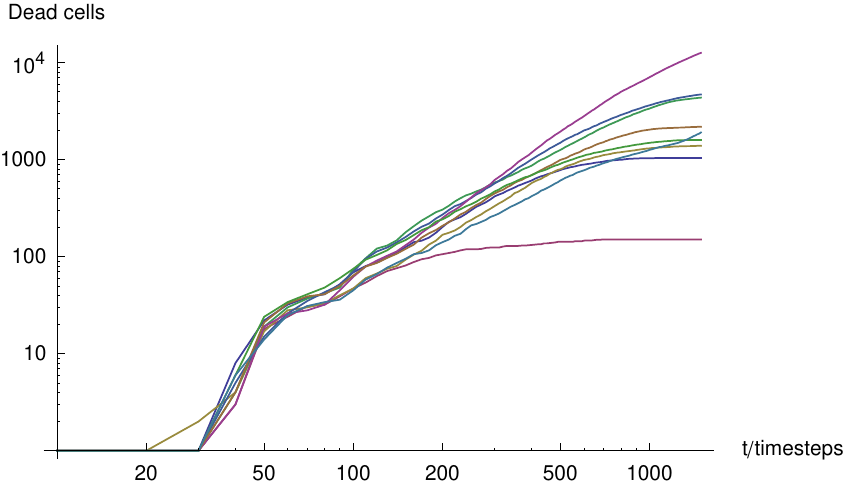}}
 \caption{The number of dead cells in plaques with identical parameter sets, but growing in the presence of interferon with various secretion rates and effect on blocking infection strengths.\label{fig:interferoneffect}}
 \end{figure}

\end{document}